\documentclass[journal]{IEEEtran}
\usepackage{amsmath,amssymb,url,amsfonts,booktabs,multirow,cite,subfigure,graphicx}
\usepackage[lined,ruled,commentsnumbered]{algorithm2e}
\allowdisplaybreaks

\begin{document}
\title{Manifold Embedded Knowledge Transfer\\ for Brain-Computer Interfaces}
\author{
    Wen Zhang and Dongrui Wu
        \thanks{
        W.~Zhang and D.~Wu are with the Ministry of Education Key Laboratory of Image Processing and Intelligent Control, School of Artificial Intelligence and Automation, Huazhong University of Science and Technology, Wuhan 430074, China. Email: wenz@hust.edu.cn, drwu@hust.edu.cn.
    }
    \thanks{
        Dongrui~Wu is the corresponding author.
    }}

\maketitle

\begin{abstract}
Transfer learning makes use of data or knowledge in one problem to help solve a different, yet related, problem. It is particularly useful in brain-computer interfaces (BCIs), for coping with variations among different subjects and/or tasks. This paper considers offline unsupervised cross-subject electroencephalogram (EEG) classification, i.e., we have labeled EEG trials from one or more source subjects, but only unlabeled EEG trials from the target subject. We propose a novel manifold embedded knowledge transfer (MEKT) approach, which first aligns the covariance matrices of the EEG trials in the Riemannian manifold, extracts features in the tangent space, and then performs domain adaptation by minimizing the joint probability distribution shift between the source and the target domains, while preserving their geometric structures. MEKT can cope with one or multiple source domains, and can be computed efficiently. We also propose a domain transferability estimation (DTE) approach to identify the most beneficial source domains, in case there are a large number of source domains. Experiments on four EEG datasets from two different BCI paradigms demonstrated that MEKT outperformed several state-of-the-art transfer learning approaches, and DTE can reduce more than half of the computational cost when the number of source subjects is large, with little sacrifice of classification accuracy.
\end{abstract}


\section{Introduction}

A brain-computer interface (BCI) provides a direct communication pathway between a user's brain and a computer \cite{wolpaw2002,bci2013book}. Electroencephalogram (EEG), a multi-channel time-series, is the most frequently used BCI input signal. There are three common paradigms in EEG-based BCIs: motor imagery (MI) \cite{He2015}, event-related potentials (ERPs) \cite{drwuTHMS2017}, and steady-state visual evoked potentials \cite{bci2013book}. The first two are the focus of this paper.

In MI tasks, the user needs to imagine the movements of his/her body parts, which causes modulations of brain rhythms in the involved cortical areas. In ERP tasks, the user is stimulated by a majority of non-target stimuli and a few target stimuli; a special ERP pattern appears in the EEG response after the user perceives a target stimulus. EEG-based BCI systems have been widely used to help people with disabilities, and also the able-bodied \cite{wolpaw2002}.

A standard EEG signal analysis pipeline consists of temporal (band-pass) filtering, spatial filtering, and classification \cite{lotte2018review}. Spatial filters such as common spatial patterns (CSP) \cite{koles1990spatial} are widely used to enhance the signal-to-noise ratio. Recently, there is a trend to utilize the covariance matrices of EEG trials, which are symmetric positive definite (SPD) and can be viewed as points on a Riemannian manifold, in EEG signal analysis \cite{arsigny2005fast,barachant2012,yger2017rieman}. For MI tasks, the discriminative information is mainly spatial, and can be directly encoded in the covariance matrices. On the contrary, the main discriminative information of ERP trials is temporal. A novel approach was proposed in \cite{korczowski2015single} to augment each EEG trial by the mean of all target trials that contain the ERP, and then their covariance matrices are computed. However, Riemannian space based approaches are computationally expensive, and not compatible with Euclidean space machine learning approaches.

A major challenge in BCIs is that different users have different neural responses to the same stimulus, and even the same user can have different neural responses to the same stimulus at different time/locations. Besides, when calibrating the BCI system, acquiring a large number of subject-specific labeled training examples for each new subject is time-consuming and expensive. Transfer learning \cite{pan2011domain,sun2016return,Gong2012,long2013transfer,zhang2017joint}, which uses data/information from one or more source domains to help the learning in a target domain, can be used to address these problems. Some representative applications of transfer learning in BCIs can be found in \cite{wu2013collaborative,drwuTNSRE2016,jayaram2016transfer,kang2009composite,lotte2010learning,jin2018adaptive}. Many researchers \cite{kang2009composite,lotte2010learning,jin2018adaptive} attempted to seek a set of subject-invariant CSP filters to increase the signal-to-noise ratio. Another pipeline is Riemannian geometry based. Zanini \emph{et al.} \cite{zanini2018transfer} proposed a Riemannian alignment (RA) framework to align the EEG covariance matrices from different subjects. He and Wu \cite{drwuEA2020} extended RA to Euclidean alignment (EA) in the Euclidean space, so that any Euclidean space classifier can be used after it.

\begin{figure*}[ht]  \centering
\includegraphics[width=0.75\textwidth,clip]{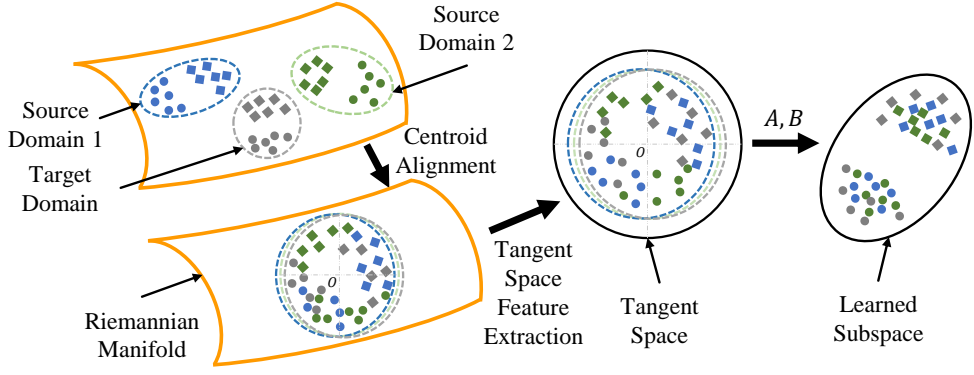}
\caption{Illustration of our proposed MEKT. Squares and circles represent examples from different classes. Different colors represent different domains. All domains are first aligned on the Riemannian manifold, and then mapped onto a tangent space. $A$ and $B$ are projection matrices of the source and the target domains, respectively.}
\label{diagram}
\end{figure*}

To utilize the excellent properties of the Riemannian geometry and avoid its high computational cost, as well as to leverage knowledge learned from the source subjects, this paper proposes a manifold embedded knowledge transfer (MEKT) framework, which first aligns the covariance matrices of the EEG trials in the Riemannian manifold, then performs domain adaptation in the tangent space by minimizing the joint probability distribution shift between the source and the target domains, while preserving their geometric structures, as illustrated in Fig.~1. Additionally, we propose a domain transferability estimation (DTE) approach to select the most beneficial subjects in multi-source transfer learning. Experiments on four datasets from two different BCI paradigms (MI and ERP) verified the effectiveness of MEKT and DTE.

The remainder of this paper is organized as follows: Section~II introduces related work on spatial filters, Riemannian geometry, tangent space mapping, RA, EA, and subspace adaptation. Section~III describes the details of the proposed MEKT and DTE. Section~IV presents experiments to compare the performance of MEKT with several state-of-the-art data alignment and transfer learning approaches. Finally, Section~V draws conclusions.

\section{Related Work}

This section introduces background knowledge on spatial filters, Riemannian geometry, tangent space mapping, RA, EA, and subspace adaptation, which will  be used in the next section.

\subsection{Spatial Filters}

Spatial filtering can be viewed as a data-driven dimensionality reduction approach that promotes the variance difference between two conditions \cite{barachant2013}. It is common in MI-based BCIs to use CSP filters \cite{Ramoser2000} to simultaneously diagonalize the two intra-class covariance matrices.

Consider a binary classification problem. Let $(X_i, y_i)$ be the $i$th labeled training example, where $X_i\in\mathbb{R}^{c\times t}$, in which $c$ is the number of EEG channels, and $t$ the number of time domain samples. For Class $k$ ($k=0,1)$, CSP finds a spatial filter matrix $W_k^*\in\mathbb{R}^{c\times f}$, where $f$ is the number of spatial filters, to maximize the variance difference between Class~$k$ and Class~$1-k$:
\begin{align}
W_k^*=\arg\max\limits_{W\in \mathbb{R}^{c\times f}}\frac{\mathrm{tr}(W^{\top}\bar{\Sigma}_k W)}
{\mathrm{tr}[W^{\top}\bar{\Sigma}_{1-k}W]}, \label{eq:CSP}
\end{align}
where $\bar{\Sigma}_k$ is the mean covariance matrix of all EEG trials in Class~$k$, and $\mathrm{tr}$ is the trace of a matrix. The solution $W_k^*$ is the concatenation of the $f$ leading eigenvectors of the matrix $(\bar{\Sigma}_{1-k})^{-1}\bar{\Sigma}_k$.

Finally, we concatenate the $2f$ spatial filters from both classes to obtain the complete CSP filters:
    \begin{align}
    W^*=[W_0^*, W_1^*]\in\mathbb{R}^{c\times 2f}  \label{eq:W}
    \end{align}
and compute the spatially filtered $X_i$ by:
\begin{align}
X_i'=(W^*)^{\top}X_i \in\mathbb{R}^{2f\times t}
\end{align}

The log-variances of the filtered trial $X'$ can be extracted:
\begin{align}
\mathbf{x}=\log \left(\frac {\operatorname{diag}(X'X'^{\top})} {\operatorname{tr} (X'X'^{\top})} \right) \label{eq:CSP2}
\end{align}
and used as input features in classification.

\subsection{Riemannian Geometry}

All SPD matrices $P\in\mathbb{R}^{c\times c}$ form a differentiable Riemannian manifold. Riemannian geometry is used to manipulate them. Some basic definitions are provided below.

The Riemannian distance between two SPD matrices $P_1$ and $P_2$ is:
\begin{align}
\delta \left(P_1, P_2\right)=\left\|\log \left(P_1^{-1} P_2\right)\right\|_F,
\end{align}
where $\|\cdot\|_{F}$ is the Frobenius norm, and $\log$ donates the logarithm of the eigenvalues of $P_1^{-1} P_2$.

The Riemannian mean of $\{P_i\}_{i=1}^n$ is:
\begin{align}
M_R=\arg\min_{P}\sum_{i=1}^n\delta^2(P,P_i), \label{eq:MR}
\end{align}

The Euclidean mean is:
\begin{align}
M_E=\frac{1}{n}\sum_{i=1}^n P_i, \label{eq:ME}
\end{align}

The Log-Euclidean mean \cite{arsigny2005fast} is:
\begin{align}
M_L=\exp \left(\sum_{i=1}^{n} w_{i} \log P_{i}\right), \label{eq:ML}
\end{align}
where $w_i$ is usually set to $\frac 1 n$.

\subsection{Tangent Space Mapping}

Tangent space mapping is also known as the logarithmic mapping, which maps a Riemannian space SPD matrix $P_i$ to a Euclidean tangent space vector $\mathbf{x}_i$ around an SPD matrix $M$, which is usually the Riemannian or Euclidean mean:
\begin{align}
\mathbf{x}_i=\mathrm{upper}\left(\log_M\left(M_{\text{ref}} P_i M_{\text{ref}}\right)\right), \label{eq:ui}
\end{align}
where $\mathrm{upper}$ takes the upper triangular part of a $c\times c$ SPD matrix and forms a vector $\mathbf{x}_i\in\mathbb{R}^{1\times c(c+1)/2}$, and $M_{\text{ref}}$ is a reference matrix. To obtain a tangent space locally homomorphic to the manifold, $M_{\text{ref}}=M^{-1/2}$ is needed \cite{barachant2013}.

Congruent transform and congruence invariance \cite{bhatia2009spd} are two important properties in the Riemannian space:
\begin{align}
\mathcal{M}\left(F P_1 F, F P_2 F\right)&=F \cdot \mathcal{M}(P_1, P_2)\cdot F, \\
\delta\left(G^{\top} P_1 G, G^{\top} P_2 G\right)&=\delta\left(P_1, P_2 \right), \label{eq:delta}
\end{align}
where $\mathcal{M}$ is the Euclidean or Riemannian mean operation, $F$ is a nonsingular square matrix, and $G\in\mathbb{R}^{c\times c}$ is an invertible symmetric matrix. (\ref{eq:delta}) suggests that the Riemannian distance between two SPD matrices does not change, if both are left and right multiplied by an invertible symmetric matrix.

\subsection{Riemannian Alignment (RA)}

RA \cite{zanini2018transfer} first computes the covariance matrices of some resting (or non-target) trials, $\{P_i\}^n_{i=1}$, in which the subject is not performing any task (or not performing the target task), and then the Riemannian mean $M_R$ of these matrices, which is used as the reference matrix to reduce the inter-session or inter-subject variations, by the following transformation:
\begin{align}
P_i'=M_R^{-1/2} P_i M_R^{-1/2}, \label{eq:ra}
\end{align}
where $P_i$ is the covariance matrix of the $i$-th trial, and $P_i'$ the corresponding aligned covariance matrix. Then, all $P_i'$ can be classified by a minimum distance to mean (MDM) classifier \cite{barachant2012}.

\subsection{Euclidean Alignment (EA)}

Although RA-MDM has demonstrated promising performance, it still has some limitations \cite{drwuEA2020}: 1) it processes the covariance matrices in the Riemannian space, whereas there are very few Riemannian space classifiers; 2) it computes the reference matrix from the non-target stimuli in ERP-based BCIs, which requires some labeled trials from the new subject.

EA \cite{drwuEA2020} extends RA and solves the above problems by transforming an EEG trial $X_i$ in the Euclidean space:
\begin{align}
X_i'=M_E^{-1/2} X_i,
\end{align}
where $M_E$ is the Euclidean mean of the covariance matrices of all EEG trials, computed by (\ref{eq:ME}).

However, EA only considers the marginal probability distribution shift, and works best when the number of EEG channels is small. When there are a large number of channels, computing $M_E^{-1/2}$ may be numerically unstable.

\subsection{Subspace Adaptation}

Tangent space vectors usually have very high dimensionality, so they cannot be used easily in transfer learning. An intuitive approach is to align them in a lower dimensional subspace. Pan \emph{et al.} \cite{pan2011domain} proposed transfer component analysis (TCA) to learn the transferable components across domains in a reproducible kernel Hilbert space using maximum mean discrepancy (MMD) \cite{gretton2012kernel}. Joint distribution adaptation (JDA) \cite{long2013transfer} improves TCA by considering the conditional distribution shift using pseudo label refinement. Joint geometrical and statistical alignment (JGSA) \cite{zhang2017joint} further improves JDA by adding two regularization terms, which minimize the within-class scatter matrix and maximize the between-class scatter matrix.

\section{Manifold Embedded Knowledge Transfer (MEKT)}

This section proposes the MEKT approach. Its goal is to use one or multiple source subjects' data to help the target subject, given that they have the same feature space and label space. For the ease of illustration, we focus on a single source domain first.

Assume the source domain has $n_S$ labeled instances $\{(X_{S,i},y_{S,i})\}_{i=1}^{n_S}$, where $X_{S,i} \in \mathbb{R}^{c\times t}$ is the $i$-th feature matrix, and $y_{S,i}\in\{1, ..., l\}$ is the corresponding label, in which $c$, $t$ and $l$ denote the number of channels, time domain samples, and classes, respectively. Let $\mathbf{y}_S=[y_{S,1}; \cdots; y_{S,n_S}]\in \mathbb{R}^{n_S\times 1}$ be the label vector of the source domain. Assume also the target domain has $n_T$ unlabeled feature matrices $\{X_{T,i}\}_{i=1}^{n_T}$, where $X_{T,i} \in \mathbb{R}^{c\times t}$.

MEKT consists of the following three steps:
\begin{enumerate}
\item \emph{Covariance matrix centroid alignment (CA)}: Align the centroids of the covariance matrices of $\{X_{S,i}\}_{i=1}^{n_S}$ and $\{X_{T,i}\}_{i=1}^{n_T}$, so that their marginal probability distributions are close.
\item \emph{Tangent space feature extraction}: Map the aligned covariance matrices to a tangent space feature matrix $X_S\in\mathbb{R}^{d \times n_S}$, and $X_T\in\mathbb{R}^{d \times n_T}$, where $d=c(c + 1)/2$ is the dimensionality of the tangent space features.
\item \emph{Mapping matrices identification}: Find projection matrices $A\in \mathbb{R}^{d \times p}$ and $B\in \mathbb{R}^{d \times p}$, where $p\ll d$ is the dimensionality of a shared subspace, such that $A^{\top}X_S$ and $B^{\top}X_T$ are close.
\end{enumerate}
After MEKT, a classifier can be trained on $(A^{\top}X_S, \mathbf{y}_S)$ and applied to $B^{\top}X_T$ to obtain the target pseudo labels $\hat {\mathbf{y}}_T$.

Next, we describe the details of the above three steps.

\subsection{Covariance Matrix Centroid Alignment (CA)} \label{sect:CA}

CA serves as a preprocessing step to reduce the marginal probability distribution shift of different domains, and enables transfer from multiple source domains.

Let $P_{S,i}=X_{S,i}X_{S,i}^{\top}$ be the $i$-th covariance matrix in the source domain, and $M_{\text{ref}}=M^{-1/2}$, where $M$ can be the Riemannian mean in (\ref{eq:MR}), the Euclidean mean in (\ref{eq:ME}), or the Log-Euclidean mean in (\ref{eq:ML}). Then, we align the covariance matrices by
\begin{align}
P_{S,i}'=M_{\text{ref}} P_{S,i} M_{\text{ref}},\qquad i=1,...,n_S \label{eq:TSM1}
\end{align}

Likewise, we can obtain the aligned covariance matrices $\{P_{T,i}'\}_{i=1}^{n_T}$ of the target domain.

CA has two desirable properties:
\begin{enumerate}
\item{\emph{Marginal probability distribution shift minimization}}.  From the properties of congruent transform and congruence invariance, we have
\begin{align}
\begin{split}
\mathcal M &(M_{\text{ref}}^{\top}P_1M_{\text{ref}}, ...,M_{\text{ref}}^{\top}P_{n_S}M_{\text{ref}})\\
& = M_{\text{ref}}^{\top} \mathcal M(P_1, ...,P_{n_S}) M_{\text{ref}} = M_{\text{ref}}^{\top} M M_{\text{ref}}=I, \label{eq:M}
\end{split}
\end{align}
i.e., if we choose $M$ as the Riemannian (or Euclidean) mean, then different domains' geometric (or arithmetic) centers all equal the identity matrix. Therefore, the marginal distributions of the source and the target domains are brought closer on the manifold.

\item{\emph{EEG trial whitening}}. In the following, we show that each aligned covariance matrix is approximately an identity matrix after CA.

If we decompose the reference matrix as $M_{\text{ref}}=\begin{bmatrix}\mathbf{w}_1, \dots, \mathbf{w}_c\end{bmatrix}$, then the $(m,n)$-th element of $P_{S,i}'$ is:
\begin{align}
P_{S,i}'(m,n)=\mathbf{w}_m^{\top} P_{S,i} \mathbf{w}_n, \label{eq:Pmn}
\end{align}

From (\ref{eq:M}) we have
\begin{align}
\mathbf{w}_m^{\top} \mathcal M(P_1, ...,P_{n_S}) \mathbf{w}_n=\left\{\begin{array}{cc}
                                                                  1, & m=n \\
                                                                  0, & m\neq n
                                                                \end{array}\right..
\end{align}
The above equation holds no matter whether $\mathcal M$ is the Riemannian mean, or the Euclidean mean.

For CA using the Euclidean mean, the average of the $m$-th diagonal element of $\{P_{S,i}'\}_{i=1}^{n_S}$ is
\begin{align}
\frac 1 {n_S}\sum_{i=1}^{n_S}P_{S,i}'(m,m) = \mathbf{w}_m^{\top} \mathcal M(P_1, ...,P_{n_S}) \mathbf{w}_m = 1, \label{eq:mean}
\end{align}
Meanwhile, for each diagonal element, we have $P_{S,i}'(m,m)=\| X_{S,i}^{\top} \mathbf{w}_m \|_2^2>0$, therefore the diagonal elements of $P_{S,i}'$ are around 1. Similarly, the off-diagonal elements of $P_{S,i}'$ are around 0. Thus, $P_{S,i}'$ is approximatively an identify matrix, i.e., the aligned EEG trials are approximated whitened.

CA with the Riemannian mean is an iterative process initialized by the Euclidean mean. CA with the Log-Euclidean mean is an approximation of CA with the Riemannian mean, with reduced computational cost \cite{arsigny2005fast}. So, (\ref{eq:mean}) also holds approximately for these two means.

This whitening effect will also be experimentally demonstrated in Section~\ref{sect:vis}.
\end{enumerate}

\subsection{Tangent Space Feature Extraction}

After covariance matrix CA, we map each covariance matrix to a tangent space feature vector in $\mathbb{R}^{d\times 1}$:
\begin{align}
\mathbf{x}_{S,i}&=\mathrm{upper}\left(\log\left(P_{S,i}' \right)\right),\qquad i=1,...,n_S \label{eq:TSM2}\\
\mathbf{x}_{T,i}&=\mathrm{upper}\left(\log\left(P_{T,i}' \right)\right),\qquad i=1,...,n_T \label{eq:TSM2T}
\end{align}
Note that this is different from the original tangent space mapping in (\ref{eq:ui}), in that (\ref{eq:ui}) uses the same reference matrix $M_{\text{ref}}$ for all subjects, whereas our approach uses a subject-specific $M_{\text{ref}}$ for each different subject.

Next, we form new feature matrices $X_S=[\mathbf{x}_{S,i},...,\mathbf{x}_{S,n_S}]$ and $X_T=[\mathbf{x}_{T,i}, ...,\mathbf{x}_{T,n_T}]$.

\subsection{Mapping Matrices Identification}

CA does not reduce the conditional probability distribution discrepancies. We next find projection matrices $A, B\in \mathbb{R}^{d \times d'}$, which map $X_S$ and $X_T$ to lower dimensional matrices $A^{\top}X_S$ and $B^{\top}X_T$, with the following desirable properties:

\begin{enumerate}
\item{\emph{Joint probability distribution shift minimization.}} In traditional domain adaptation \cite{pan2011domain,long2013transfer}, MMD is frequently used to reduce the marginal and conditional probability distribution discrepancies between the source and the target domains, i.e.,
\begin{align}
\begin{split}
\mathcal{D}_{S,T} &\approx \mathcal{D}\left(Q\left(X_S\right), Q\left(X_T\right)\right) \\
&\quad +\mathcal{D}\left(Q\left(\mathbf{y}_S | X_S\right), Q\left(\hat {\mathbf{y}}_T | X_T\right)\right) \\
&=\left\|\frac{1}{n_S} \sum_{i=1}^{n_S} A^{\top} \mathbf{x}_{S, i}-\frac{1}{n_T} \sum_{j=1}^{n_T} B^{\top} \mathbf{x}_{T, j}\right\|_{F}^{2} \\
&\quad + \sum_{k=1}^l {\left \| \frac 1 {n_S^k}\sum_{i=1}^{n_S^k}A^{\top}\mathbf{x}_{S,i}^k - \frac 1 {n_T^k}\sum_{j=1}^{n_T^k}B^{\top}\mathbf{x}_{T,j}^k \right \| _F^2}, \label{eq:MMD}
\end{split}
\end{align}
where $\mathbf{x}_{S,i}^k$ and $\mathbf{x}_{T,j}^k$ are the tangent space vectors in the $k$-th ($k=1,...,l$) class of the source domain and the target domain, respectively, and $n_S^k$ and $n_T^k$ are the number of examples in the $k$-th class of the source domain and the target domain, respectively.

Next, we propose a new measure, joint probability MMD, to quantify the probability distribution shift between the source and the target domains, by considering the joint probability directly, instead of the marginal and the conditional probabilities separately.

Then, the joint probability MMD between the source and the target domains is:
\begin{align}
\begin{split}
\mathcal{D}'_{S,T} &= \mathcal{D}\left(Q\left(X_S,\mathbf{y}_S\right), Q\left(X_T,\hat {\mathbf{y}}_T\right)\right) \\
& = \mathcal{D}\left(Q\left(X_S|\mathbf{y}_S\right)Q(\mathbf{y}_S), Q\left(X_T|\hat {\mathbf{y}}_T\right)Q(\hat {\mathbf{y}}_T)\right)\\
& \approx \sum_{k=1}^l {\left \| \frac {P(\mathbf{y}_S^k)} {n_S^k}\sum_{i=1}^{n_S^k}A^{\top}\mathbf{x}_{S,i}^k - \frac {P(\hat{\mathbf{y}}_T^k)} {n_T^k}\sum_{j=1}^{n_T^k}B^{\top}\mathbf{x}_{T,j}^k \right \| _F^2}\\
& = \sum_{k=1}^l {\left \| \frac 1 {n_S}\sum_{i=1}^{n_S^k}A^{\top}\mathbf{x}_{S,i}^k - \frac 1 {n_T}\sum_{j=1}^{n_T^k}B^{\top}\mathbf{x}_{T,j}^k \right \| _F^2}, \label{eq:JPMMD}
\end{split}
\end{align}

Let the one-hot encoding matrix of the source domain label vector\footnote{For example, for binary classification with two classes 1 and 2, if $\mathbf{y}_S=\left[\begin{array}{c}
                                        2 \\
                                        1  \\
                                        2
                                      \end{array}\right]$, then $Y_S=\left[\begin{array}{cc}
                                        0 & 1 \\
                                        1 & 0 \\
                                        0 & 1
                                      \end{array}\right]$.}
$\mathbf{y}_S$ be $Y_S$, and the one-hot encoding matrix of the \emph{predicted} target label vector $\hat {\mathbf{y}}_T$ be $\hat{Y}_T$. (\ref{eq:JPMMD}) can be simplified as
\begin{align}
\mathcal{D}'_{S,T} = \left \| N_S^{\top}X_S^{\top}A-N_T^{\top}X_T^{\top}B \right \|_F^2, \label{eq:fJPMMD}
\end{align}
where
\begin{align}
N_S=\frac {Y_S} {n_S},\qquad N_T=\frac {\hat Y_T} {n_T}. \label{eq:NT}
\end{align}
The joint probability MMD is based on the joint probability rather than the conditional probability, which in theory can handle more probability distribution shifts.

\item{\emph{Source domain discriminability.}} During subspace mapping, the discriminating ability of the source domain can be preserved by:
\begin{align}
\min_A \ \operatorname{tr}(A^{\top}S_wA)\qquad s.t.\ A^{\top}S_bA=I, \label{eq:sb}
\end{align}
where $S_w=\sum_{k=1}^l \sum_{i=1}^{n_S^k} (\mathbf{x}_{S,i}^k)^{\top} \mathbf{x}_{S,i}^k h_k$ is the within-class scatter matrix, in which $h_k=1-\frac 1 {n_S^k}$, and $S_b=\sum_{k=1}^l n_k \left(\bar{m}_k-\bar{m} \right)\left(\bar{m}_k-\bar{m} \right)^{\top} $ is the between-class scatter matrix, in which $\bar{m}_k$ is the mean of samples from Class~$k$, and $\bar{m}$ is the mean of all samples.

\item{\emph{Target domain locality preservation.}} We also introduce a graph-based regularization term to preserve the local structure in the target domain. Under the manifold assumption \cite{belkin2004semi}, if two samples $\mathbf{x}_{T,i}$ and $\mathbf{x}_{T,j}$ are close in the original target domain, then they should also be close in the projected subspace.

Let $S\in\mathbb{R}^{n_T\times n_T}$ be a similarity matrix:
\begin{align}
S_{i j}=\begin{cases}e^{\frac{-\left \| \mathbf{x}_{T,i} - \mathbf{x}_{T,j} \right \|_2^2}{2\sigma^2} }, & \text{if } \mathbf{x}_{T,i} \in N_p(\mathbf{x}_{T,j})\\
&\text{ or } \mathbf{x}_{T,j} \in N_p(\mathbf{x}_{T,i}) \\
0,& \text{otherwise}\end{cases},
\end{align}
where $N_p(\mathbf{x}_{T,i})$ is the set of the $p$-nearest neighbors of $\mathbf{x}_{T,i}$, and $\sigma$ is a scaling parameter, which usually equals 1 \cite{cai2005document}.

Using the normalized graph Laplacian matrix $L=I-D^{-1/2}SD^{-1/2}$, where $D$ is a diagonal matrix with $D_{ii}=\sum_{j=1}^{n_T} S_{ij}$, graph regularization is expressed as:
\begin{align}
\sum_{i,j=1}^{n_T}\|B^{\top}\mathbf{x}_{T,i}-B^{\top}\mathbf{x}_{T,j}\|_2^2S_{ij}=\operatorname{tr}(B^{\top}X_TLX_T^{\top}B),
\end{align}

To remove the scaling effect, we add a constraint on the target embedding \cite{belkin2003laplacian}:
\begin{align}
\min_{B}\operatorname{tr}(B^{\top}X_TLX_T^{\top}B) \quad s.t.\ B^{\top}X_THX_T^{\top}B=I,
\end{align}
where $H=I-\frac{1}{n_T}\mathbf{1}_{n_T} $ is the centering matrix, in which $\mathbf{1}_{n_T}\in \mathbb{R}^{n_T\times n_T}$ is an all-one matrix.

\item{\emph{Parameter transfer and regularization.}}  Since the source and the target domains have the same feature space, and CA has brought their probability distributions closer, we want the projection matrix $B$ to be similar to the projection matrix $A$ learned in the source domain. Additionally, for better generalization performance, we want to ensure that $A$ and $B$ do not include extreme values. Thus, we have the following constraints on the projection matrices:
\begin{align}
\min_{A,B}\left(\|B-A\|_F^2+\|B\|_F^2\right).
\end{align}
\end{enumerate}

\subsection{The Overall Loss Function of MEKT}

Integrating all regularization and constraints above, the formulation of MEKT is:
\begin{align}
\begin{split}
\min_{A,B}& \ \alpha \operatorname{tr}(A^{\top}S_wA) + \beta \operatorname{tr}(B^{\top}X_TLX_T^{\top}B) + \mathcal{D}'_{S,T} \\
&+\rho(\|B-A\|_F^2+\|B\|_F^2)\\
&s.t.\quad B^{\top}X_THX_T^{\top}B=I,\quad A^{\top}S_bA=I \label{eq:MEKT}
\end{split}
\end{align}
where $\alpha$, $\beta$ and $\rho$ are trade-off parameters to balance the importance of the source domain discriminability, the target domain locality, and the parameter regularization, respectively.

Let $W=[A; B]$. Then, the Lagrange function is
\begin{align}
\begin{split}
\mathcal{J}=\operatorname{tr} \left( W^{\top}(\alpha P+\beta L +\rho U +R) W+ \eta(I-W^{\top}V W)\right) \label{eq:J}
\end{split}
\end{align}
where
\begin{align}
P&=\left[ \begin{matrix} S_w & \mathbf{0}\\ \mathbf{0} & \mathbf{0} \end{matrix} \right],\quad
L=\left[ \begin{matrix} \mathbf{0} & \mathbf{0} \\ \mathbf{0} & X_TLX_T^{\top} \end{matrix} \right], \label{eq:PL}\\
U&=\left[ \begin{matrix} I & -I\\ -I & 2I \end{matrix} \right],\quad
V=\left[ \begin{matrix} S_b & \mathbf{0}\\ \mathbf{0} & X_THX_T^{\top} \end{matrix} \right], \label{eq:UV} \\
R&=\left[ \begin{matrix} X_SN_SN_S^{\top}X_S^{\top} & -X_SN_SN_T^{\top}X_T^{\top} \\ -X_TN_TN_S^{\top}X_S^{\top} & X_TN_TN_T^{\top}X_T^{\top} \end{matrix} \right], \label{eq:R}
\end{align}

Setting the derivative $\nabla_W \mathcal{J}=\mathbf{0}$, we have
\begin{align}
(\alpha P+\beta L +\rho U +R) W=\eta V W \label{eq:W}
\end{align}
(\ref{eq:W}) can be solved by generalized eigen-decomposition, and $W$ consists of the $p$ trailing eigenvectors. Since $\hat{Y}_T$ is needed in $N_T$ [see (\ref{eq:NT})], and hence $R$, we use a general expectation-maximization like pseudo label refinement procedure \cite{long2013transfer} to refine the estimation, as shown in Algorithm~\ref{alg:1}.

\begin{algorithm}[htpb]
\caption{Manifold Embedded Knowledge Transfer (MEKT)}
\label{alg:1}
\KwIn {$n_S$ source domain samples $\{(X_{S,i},y_{S,i})\}_{i=1}^{n_S}$, where $X_{S,i}\in\mathbb{R}^{c\times t}$ and $y_{S,i}\in\{1, ..., l\}$\;
\hspace*{10mm}$n_T$ target domain feature matrices $\{X_{T,i}\}_{i=1}^{n_T}$, where $X_{T,i}\in\mathbb{R}^{c\times t}$\;
\hspace*{10mm}Number of iterations $N$\;
\hspace*{10mm}Weights $\alpha$, $\beta$, $\rho$\;
\hspace*{10mm}Dimensionality of the shared subspace, $p$.}
\KwOut {$\hat {\mathbf{y}}_T\in\mathbb{R}^{n_T\times 1}$, the labels for $\{X_{T,i}\}_{i=1}^{n_T}$.}
Calculate the covariance matrices $\{P_{S,i}\}_{i=1}^{n_S}$ and their mean matrix $M$ in the source domain, using (\ref{eq:MR}), (\ref{eq:ME}), or (\ref{eq:ML})\;
Calculate $\{P_{S,i}'\}_{i=1}^{n_S}$ using (\ref{eq:TSM1})\;
Map each $P_{S,i}'$ to a tangent space feature vector $\mathbf{x}_{S,i}\in\mathbb{R}^{d \times 1}$ using (\ref{eq:TSM2}) ($d=c(c+1)/2$)\;
Repeat the above procedure to get $\mathbf{x}_{T,i}\in\mathbb{R}^{d \times 1}$ using (\ref{eq:TSM2T})\;
Form $X_S=[\mathbf{x}_{S,1}, ...,\mathbf{x}_{S,n_S}]$ and $X_T=[\mathbf{x}_{T,1}, ...,\mathbf{x}_{T,n_T}]$\;
Construct $P$, $L$, $U$, $V$ and $R$ in (\ref{eq:PL})-(\ref{eq:R})\;
\For{$n=1,...,N$}
{
    Solve (\ref{eq:W}), and construct $W\in\mathbb{R}^{2d \times p}$ as the $p$ trailing eigenvectors\;
    Construct $A$ as the first $d$ rows in $W$, and $B$ as the last $d$ rows\;
    Train a classifier $f$ on $(A^{\top}X_S,\mathbf{y}_S)$\ and apply it to $B^{\top}X_T$ to update $\hat {\mathbf{y}}_T$\;
    Update $R$ in (\ref{eq:R}).
}
\Return $\hat {\mathbf{y}}_T$.
\end{algorithm}

Note that for the clarity of explanation, Algorithm~1 only considers one source domain. When there are multiple source domains, we perform CA and compute the tangent space feature vectors $X_S^{(i)}\in \mathbb{R}^{d\times n_S^{(i)}}$ for each source domain separately, and then assemble their feature vectors into a single source domain feature matrix $X_S=[X_S^{(1)}, ..., X_S^{(z)}]\in \mathbb{R}^{d\times n^{\ast}}$, where $n_S^{(i)}$ is the number of trails in the $i$-th source domain, $z$ is the number of source domains, and $n^{\ast}=\sum_{i=1}^zn_S^{(i)}$.

\subsection{Kernelization Analysis}

Nonlinear MEKT can be achieved through kernelization in a Reproducing Kernel Hilbert Space \cite{zhang2017joint}.

Let the kernel function be $\phi: \mathbf{x} \mapsto \phi(\mathbf{x})$. Define $\Phi(X)=[\phi(\mathbf{x}_1),..,\phi(\mathbf{x}_n)] \in \mathbb{R}^{d\times n}$, where $n=n_S+n_T$. We use the Representer Theorem \cite{belkin2006manifold} $A=\Phi(X)\mathbf{A}$ and $B=\Phi(X)\mathbf{B}$ to kernelize MEKT, where $X=[X_S, X_T]$, and $\mathbf{A}\in \mathbb{R}^{n\times p}$ and $\mathbf{B}\in \mathbb{R}^{n\times p}$ are two projection matrices to be optimized.

Let $K_S=\Phi(X)^{\top}\Phi(X_S)$ and $K_T=\Phi(X)^{\top}\Phi(X_T)$. Then, all $\mathbf{x}$ are replaced by $\phi(\mathbf{x})$, $X_S$ by $\Phi(X_S)$, and $X_T$ by $\Phi(X_T)$, in the above derivations. The optimization problem becomes
\begin{align}
\begin{split}
\min_{\mathbf{A},\mathbf{B}}& \ \alpha \operatorname{tr}(\mathbf{A}^{\top}S_w\mathbf{A}) + \beta \operatorname{tr}(\mathbf{B}^{\top}K_TLK_T^{\top}\mathbf{B})\\
&+ \left \| N_S^{\top}K_S^{\top}\mathbf{A}-N_T^{\top}K_T^{\top}\mathbf{B} \right \|_F^2 +\rho(W^{\top}\mathbf{U}W)\\
&s.t.\quad \mathbf{B}^{\top}K_THK_T^{\top}\mathbf{B}=I,\quad \mathbf{A}^{\top}S_b\mathbf{A}=I \label{eq:kloss2}
\end{split}
\end{align}
where $S_w=\sum_{k=1}^l K_S^kH_S^k(K_S^k)^{\top}$, in which $K_S^k$ is the part of $K_S$ from Class~$k$ only, and $H_S^k=I-\frac{1}{n_S^k}\mathbf{1}$ the centering matrix. The Laplacian matrix $L$ is constructed in the original data space. In $S_b$, $\bar{m}_k$ is the mean of $K_S^k$, and $\bar{m}$ the mean of $K=[K_S, K_T]$. $\mathbf{U}$ is obtained by replacing $I$ in (\ref{eq:UV}) with $K$.

(\ref{eq:kloss2}) can be optimized in a similar way as (\ref{eq:MEKT}).

\subsection{Domain Transferability Estimation (DTE)}

When there are a large number of source domains, estimating domain transferability can advise which domains are more important, and also reduce the computational cost. In BCIs, DTE can be used to find subjects which have low correlations to the tasks and hence may cause negative transfer. Although source domain selection is important, it is very challenging, and hence very few publications can be found in the literature \cite{Gong2012,drwuTHMS2017,drwuTFS2017,Wei2015}.

Next, we propose an unsupervised DTE strategy.

Assume there are $z$ labeled sources domains $\mathbb{S}_i=\{X_S^{(i)}, \mathbf{y}_S^{(i)}\}_{i=1}^z$, where $X_S^{(i)}$ is the feature matrix of the $i$-th source domain, $\mathbf{y}_S^{(i)}$ is the corresponding label vector. Assume also there is a target domain $\mathbb{T}$ with unlabeled feature matrix $X_T$. Let $S_b$ be the between-class scatter matrix, similar to $S_b$ in (\ref{eq:sb}), and $S_b^{\mathbb{S}_i,\mathbb{T}}$ be the scatter matrix between the source and the target domains. We define the discriminability of the $i$-th source domain as $DIS(\mathbb{S}_i)=\|S_b^{\mathbb{S}_i}\|_1$, and the difference between the source domain and the target domain as $DIF(\mathbb{S}_i,\mathbb{T})=\|S_b^{\mathbb{S}_i,\mathbb{T}}\|_1$.

Then, the transferability of Source Domain~$\mathbb{S}_i$ is computed as:
\begin{align}
r(\mathbb{S}_i,\mathbb{T})=\frac {DIS(\mathbb{S}_i)} {DIF(\mathbb{S}_i,\mathbb{T})} \label{eq:dte}
\end{align}

We then select $z^{\ast}\in(1,z)$ source subjects with the highest $r(\mathbb{S}_i,\mathbb{T})$.

\section{Experiments}

In this section, we evaluate our method for both single-source to single-target (STS) transfers and multi-source to single-target (MTS) transfers. The code is available online\footnote{https://github.com/chamwen/MEKT}.

\subsection{Datasets}

We used two MI datasets and two ERP datasets in our experiments. Their statistics are summarized in Table~\ref{tab:data}.

\begin{table}[htbp] \centering \setlength{\tabcolsep}{1.2mm}
  \caption{Statistics of the two MI and two ERP datasets.}
    \begin{tabular}{c|ccccc} \toprule
    \multirow{2}{*}{Dataset}  & Number of & Number of & Number of & Trails per & Class- \\
      & Subjects & Channels & Time Samples & Subject & Imbalance \\ \midrule
    MI1   & 7     & 59    & 300   & 200   & No \\
    MI2   & 9     & 22    & 750   & 144   & No \\
    RSVP  & 11    & 8     & 45    & 368-565 & Yes \\
    ERN   & 16    & 56    & 260   & 340   & Yes \\ \bottomrule
    \end{tabular}   \label{tab:data}
\end{table}

For both MI datasets, a subject sat in front of a computer screen. At the beginning of a trial, a fixation cross appeared on the black screen to prompt the subject to be prepared. Shortly after, an arrow pointing to a certain direction was presented as a visual cue for a few seconds, during which the subject performed a specific MI task. Then, the visual cue disappeared, and the next trial started after a short break. EEG signal was recorded during the experiment, and used to classify which MI the user was performing. Usually, EEG shortly after the visual cue onset is highly related to the MI task.

\begin{figure}[htpb]  \centering
\includegraphics[width=\linewidth,clip]{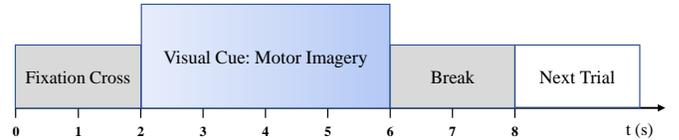}
\caption{Timing scheme of the motor imagery tasks in the first two datasets.} \label{fig:mi}
\end{figure}

For the first MI dataset\footnote{http://www.bbci.de/competition/iv/desc\_1.html.} (MI1), 59-channel EEGs were recorded at 100 Hz from seven healthy subjects, each with 100 left hand MIs and 100 right hand MIs. For the second MI dataset\footnote{http://www.bbci.de/competition/iv/desc\_2a.pdf.} (MI2), 22-channel EEGs were recorded at 250 Hz from nine heathy subjects, each with 72 left hand MIs and 72 right hand MIs. Both datasets were used for two-class classification.

The first ERP dataset\footnote{https://www.physionet.org/physiobank/database/ltrsvp/.} contained 8-channel EEG recordings from 11 healthy subjects in a rapid serial visual presentation (RSVP) experiment. The images were presented at different rates (5, 6, and 10 Hz) in three different experiments. We only used the 5 Hz version. The goal was to classify from EEG if the subject had seen a target image (with airplane) and non-target image (without airplane). The number of images for different subjects varying between 368 and 565, and the target to non-target ratio was around 1:9. The sampling rate was 2048 Hz, and the RSVP data had been band-pass filtered to 0.15-28 Hz.

The second ERP dataset\footnote{https://www.kaggle.com/c/inria-bci-challenge.} was recorded from a feedback error-related negativity (ERN) experiment \cite{margaux2012}, which was used in a Kaggle competition for two-class classification. It was collected from 26 subjects and partitioned into training set (16 subjects) and test set (10 subjects). We only used the 16 subjects in the training set as we do not have access to the test set. The average target to non-target ratio was around 1:4. The 56-channel EEG data had been downsampled to 200 Hz.

\subsection{EEG Data Preprocessing}

EEG signals from all datasets were preprocessed using the EEGLAB toolbox \cite{delorme2004eeglab}. We followed the same preprocessing procedures in \cite{drwuEA2020,zhang2019vulnerability}.

For the two MI datasets, a causal 50-order 8-30 Hz\footnote{We bandpass filtered the EEG signal to 8-30 Hz because MI is mainly indicated by the change of the mu rhythm (about 8–13 Hz) and the beta (about 14–30 Hz) rhythm \cite{Ai2018}.} finite impulse response (FIR) band-pass filter was used to remove muscle artifacts and direct current drift, and hence to obtain cleaner MI signals. Next, EEG signals between $[0.5,3.5]$ seconds after the cue onsets were extracted as trials. The RSVP signal was downsampled to 64 Hz to reduce the computational cost, and epoched to 0.7s intervals immediately after the stimulus onsets as trials. The ERN signal was bandpass filtered to 1-40 Hz, and epoched to 1.3s intervals immediately after the feedbacks (which contained the ERP associated with the user's response to the feedback event) as trials.

MI1 had 59 EEG channels, which were not easy to manipulate. Thus, we reduced the number of its tangent space features to the number of source domain samples (200), according to their $F$ values in one-way ANOVA. For the ERN dataset, we used xDAWN \cite{rivet2009xdawn} to reduce the number of channels from 56 to 6.

The dimensionalities of different input spaces are shown in Table~\ref{tab:data2}. $n_i$ is the number of samples in the $i$-th domain, and $c$ the number of selected channels for the two ERP datasets. Specifically, for RSVP, $c=8$ and $n_i$ varies from 368 to 565; for ERN, $c=6$ and $n_i=340$. Augmented covariance matrices \cite{korczowski2015single} were used in the Riemannian space for ERP, so they had dimensionality of $2c\times 2c$. Only the $c\times c$ upper right block of the augmented covariance matrix contains temporal information \cite{korczowski2015single}, so these $c^2$ elements were selected as the tangent space features.

\begin{table}[htbp] \centering
  \caption{Input space dimensionalities in different STS tasks.}
    \begin{tabular}{c|cccc} \toprule
    & MI1 & MI2 & ERP (RSVP and ERN)\\ \midrule
    Euclidean  & 6$\times$200 & 6$\times 144$ & 20$\times n_i$ \\
    Tangent    & 200$\times$200 & 253$\times 144$ & $ c^2 \times n_i$ \\
    Riemannian & 59$\times$59$\times$200 & 22$\times$22$\times 144$ & $2c \times 2c \times n_i$ \\ \bottomrule
    \end{tabular}  \label{tab:data2}
\end{table}

Next, we describe how the Euclidean space features were determined. For the two MI datasets, six log-variance features of the CSP filtered trials [see (\ref{eq:CSP2})] were used as features. For the two ERP datasets, after spatial filtering by xDAWN, we assembled each EEG trail (which is a matrix) into a vector, performed principal component analysis on all vectors from the source subjects, and extracted the scores for the first 20 principal components as features.

\subsection{Baseline Algorithms}

We compared our MEKT approaches (MEKT-R: the Riemannian mean is used as the reference matrix; MEKT-E: the Euclidean mean is used as the reference matrix; MEKT-L: the Log-Euclidean mean is used as the reference matrix) with seven state-of-the-art baseline algorithms for BCI classification. According to the feature space type, these baselines can be divided into three categories:
\begin{enumerate}
\item{\emph{Euclidean} space approaches:}
\begin{enumerate}
\item CSP-LDA (linear discriminant analysis) \cite{bishop2006pattern} for MI, and CSP-SVM (support vector machine) \cite{chang2011svm} for ERP.
\item EA-CSP-LDA for MI, and EA-xDAWN-SVM for ERP, i.e., we performed EA \cite{drwuEA2020} as a preprocessing step before spatial filtering and classification.
\end{enumerate}

\item{\emph{Riemannian} space approach:} RA-MDM \cite{zanini2018transfer} for MI, and xDAWN-RA-MDM for ERP.

\item{\emph{Tangent} space approaches, which were proposed for computer vision applications, and have not been used in BCIs before. CA was used before each of them. In each learned subspace, the sLDA classifier \cite{peck1982use} was used for MI, and SVM for ERP.}
\begin{enumerate}
\item CA (centroid alignment).
\item CA-CORAL (correlation alignment) \cite{sun2016return}.
\item CA-GFK (geodesic flow kernel) \cite{Gong2012}.
\item CA-JDA (joint distribution adaptation) \cite{long2013transfer}.
\item CA-JGSA (joint geometrical and statistical alignment) \cite{zhang2017joint}.
\end{enumerate}
\end{enumerate}

Hyper-parameters of all baselines were set according to the recommendations in their corresponding publications. For MEKT, $T=5$, $\alpha=0.01$, $\beta=0.1$, $\rho=20$, and $d=10$ were used.

\subsection{Experimental Settings}

We evaluated unsupervised STS and MTS transfers. In STS, one subject was selected as the target, and another as the source. Let $z$ be the number of subjects in a dataset. Then, there were $z(z-1)$ different STS tasks. In MTS, one subject was used as the target, and all others as the sources, so there were $z$ different MTS tasks. For example, MI1 included seven subjects, so we had $7\times6=42$ STS tasks, e.g., S2$\to$S1 (Subject 2 as the source, and Subject 1 as the target), S3$\to$S1, S4$\to$S1, S5$\to$S1, S6$\to$S1, S7$\to$S1, ... S6$\to$S7, and seven MTS tasks, e.g., \{S2, S3, S4, S5, S6, S7\}$\to$S1, $\ldots$, \{S1, S2, S3, S4, S5, S6\}$\to$S7.

The balanced classification accuracy (BCA) was used as the performance measure:
\begin{align}
BCA=\frac {1}{l} \sum_{k=1}^l \frac{tp_k}{n_k},
\end{align}
where $tp_k$ and $n_k$ are the number of true positives and the number of samples in Class~$k$, respectively.

\subsection{Visualization} \label{sect:vis}

\begin{figure}[htpb]  \centering
\includegraphics[width=\linewidth,clip]{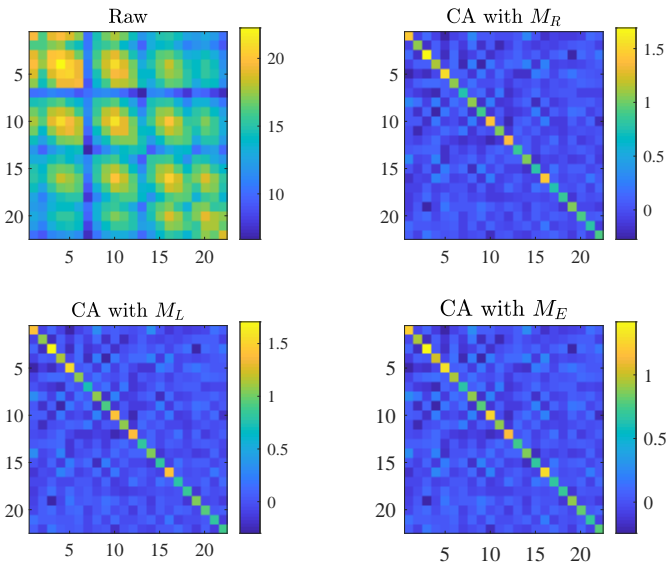}
\caption{The raw covariance matrix (Trial~1, Subject~1, MI2), and those after CA using different reference matrices.} \label{fig:ca}
\end{figure}

As explained in Section~\ref{sect:CA}, CA makes the aligned covariance matrices approximate the identity matrix, no matter whether the Riemannian mean, or the Euclidean mean, or the Log-Euclidean mean, is used as the reference matrix. To demonstrate that, Fig.~\ref{fig:ca} shows the raw covariance matrix of the first EEG trial of Subject~1 in MI2, and the aligned covariance matrices using different references. The raw covariance matrix is nowhere close to identity, but after CA, the covariance matrices are approximately identity, and hence the corresponding EEG trials are approximately whitened.

\begin{figure}[htpb]  \centering
\includegraphics[width=\linewidth,clip]{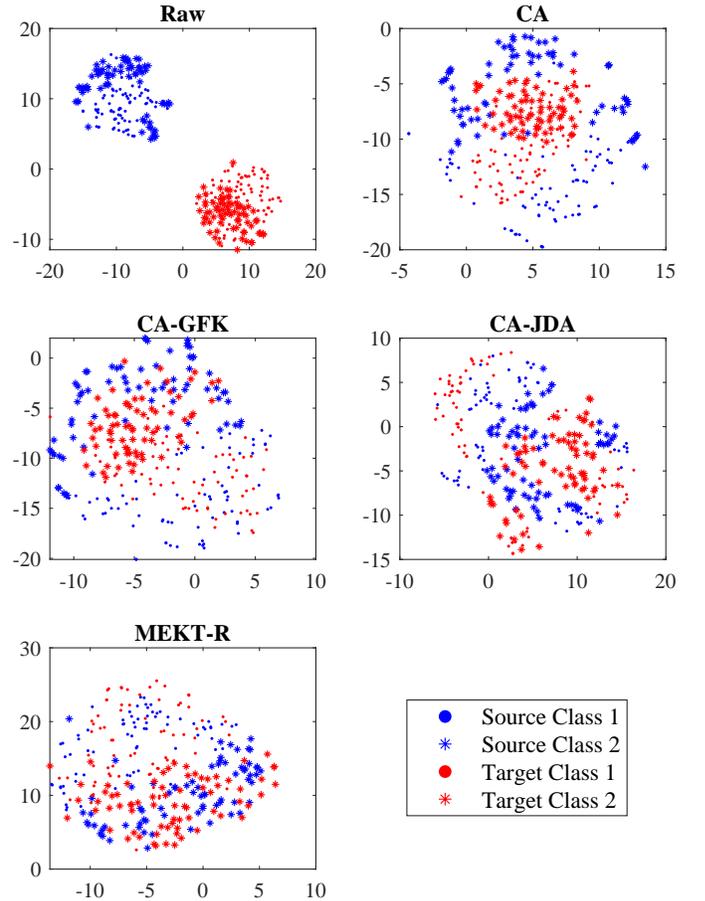}
\caption{$t$-SNE visualization of the data distributions before and after CA, and with different transfer learning approaches, when transferring Subject~2's data (source) to Subject~1 (target) in MI2.} \label{fig:view}
\end{figure}

Next, we used $t$-SNE \cite{maaten2008} to reduce the dimensionality of the EEG trials to two, and visualize if MEKT can bring the data distributions of the source and the target domains together. Fig.~\ref{fig:view} shows the results on transferring Subject~2's data to Subject~1 in MI2, before and after different data alignment approaches. Before CA, the source domain and target domain samples do not overlap at all. After CA, the two sets of samples have identical mean, but different variances. CA-GFK and CA-JDA make the variance of the source domain samples and the variance of the target domain samples approximately identical, but different classes are still not well separated. MEKT-R not only makes the overall distributions of the source domain samples and the target domain samples consistent, but also samples from the same class in the two domains close, which should benefit the classification.

\subsection{Classification Accuracies}

The means and standard deviations of the BCAs on the four datasets with STS and MTS transfers are shown in Tables~\ref{tab:bca1} and \ref{tab:bca2}, respectively. All MEKT-based approaches achieved the best (in bold) or the second best (underlined) performance in all scenarios in contrast to the baselines.

\begin{table}[htbp] \centering \setlength{\tabcolsep}{4mm}
  \caption{Mean (\%) and standard deviation (\%; in parenthesis) of the BCAs in STS transfers. For the CA-based approaches, the sLDA classifier was used for MI, and SVM for ERP.}
    \begin{tabular}{l|cc|cc|c} \toprule
          & \multicolumn{2}{c|}{MI1} & \multicolumn{2}{c|}{MI2} & Avg \\ \midrule
     CSP-LDA  & \multicolumn{2}{c|}{57.23 (10.56)} & \multicolumn{2}{c|}{58.7 (11.58)} & 57.97 \\
    EA-CSP-LDA & \multicolumn{2}{c|}{66.85 (10.56)} & \multicolumn{2}{c|}{65.00 (14.06)} & 65.93 \\
     RA-MDM & \multicolumn{2}{c|}{64.98 (10.37)} & \multicolumn{2}{c|}{66.60 (12.60)} & 65.79 \\
    CA    & \multicolumn{2}{c|}{66.17 (9.93)} & \multicolumn{2}{c|}{66.02 (13.14)} & 66.10 \\
    CA-CORAL & \multicolumn{2}{c|}{67.69 (10.68)} & \multicolumn{2}{c|}{67.26 (13.34)} & 67.48 \\
    CA-GFK & \multicolumn{2}{c|}{66.62 (10.53)} & \multicolumn{2}{c|}{65.54 (13.56)} & 66.08 \\
    CA-JDA & \multicolumn{2}{c|}{66.01 (12.55)} & \multicolumn{2}{c|}{66.59 (15.28)} & 66.30 \\
    CA-JGSA & \multicolumn{2}{c|}{65.81 (13.06)} & \multicolumn{2}{c|}{65.90 (16.73)} & 65.85 \\ \midrule
    MEKT-E & \multicolumn{2}{c|}{69.19 (12.84)} & \multicolumn{2}{c|}{68.34 (15.51)} & 68.77 \\
    MEKT-L & \multicolumn{2}{c|}{\underline{70.74} (12.28)} & \multicolumn{2}{c|}{\underline{68.56} (15.66)} & \underline{69.65} \\
    MEKT-R & \multicolumn{2}{c|}{\textbf{70.99} (12.46)} & \multicolumn{2}{c|}{\textbf{68.73} (15.73)} & \textbf{69.86} \\ \midrule
          & \multicolumn{2}{c|}{RSVP} & \multicolumn{2}{c|}{ERN} & Avg \\ \midrule
     CSP-LDA  & \multicolumn{2}{c|}{58.58 (7.98)} & \multicolumn{2}{c|}{54.34 (5.87)} & 56.46 \\
    EA-CSP-LDA & \multicolumn{2}{c|}{58.76 (7.51)} & \multicolumn{2}{c|}{55.57 (6.26)} & 57.17 \\
     RA-MDM & \multicolumn{2}{c|}{60.37 (8.05)} & \multicolumn{2}{c|}{56.22 (6.89)} & 58.30 \\
    CA    & \multicolumn{2}{c|}{58.34 (6.98)} & \multicolumn{2}{c|}{56.97 (7.06)} & 57.66 \\
    CA-CORAL & \multicolumn{2}{c|}{58.45 (6.84)} & \multicolumn{2}{c|}{57.04 (7.00)} & 57.75 \\
    CA-GFK & \multicolumn{2}{c|}{59.93 (7.61)} & \multicolumn{2}{c|}{57.24 (7.34)} & 58.59 \\
    CA-JDA & \multicolumn{2}{c|}{60.27 (7.75)} & \multicolumn{2}{c|}{57.56 (7.63)} & 58.92 \\
    CA-JGSA & \multicolumn{2}{c|}{55.23 (6.74)} & \multicolumn{2}{c|}{57.17 (7.72)} & 56.20 \\ \midrule
    MEKT-E & \multicolumn{2}{c|}{61.08 (8.59)} & \multicolumn{2}{c|}{\textbf{58.01} (7.76)} & \textbf{59.55} \\
    MEKT-L & \multicolumn{2}{c|}{\underline{61.15} (8.44)} & \multicolumn{2}{c|}{\underline{57.91} (7.74)} & 59.53 \\
    MEKT-R & \multicolumn{2}{c|}{\textbf{61.24} (8.36)} & \multicolumn{2}{c|}{57.85 (7.75)} & \textbf{59.55} \\ \bottomrule
    \end{tabular}
  \label{tab:bca1}
\end{table}

\begin{table}[htbp] \centering \setlength{\tabcolsep}{4mm}
  \caption{Mean (\%) and standard deviation (\%; in parenthesis) of the BCAs in MTS transfers.}
    \begin{tabular}{l|cc|cc|c} \toprule
          & \multicolumn{2}{c|}{MI1} & \multicolumn{2}{c|}{MI2} & Avg \\ \midrule
     CSP-LDA  & \multicolumn{2}{c|}{59.71 (12.93)} & \multicolumn{2}{c|}{67.75 (12.92)} & 63.73 \\
    EA-CSP-LDA & \multicolumn{2}{c|}{79.79 (6.57)} & \multicolumn{2}{c|}{73.53 (15.96)} & 76.66 \\
     RA-MDM & \multicolumn{2}{c|}{73.29 (9.25)} & \multicolumn{2}{c|}{72.07 (9.88)} & 72.68 \\
    CA    & \multicolumn{2}{c|}{76.29 (9.66)} & \multicolumn{2}{c|}{71.84 (13.89)} & 74.07 \\
    CA-CORAL & \multicolumn{2}{c|}{78.86 (8.73)} & \multicolumn{2}{c|}{72.38 (13.38)} & 75.62 \\
    CA-GFK & \multicolumn{2}{c|}{76.79 (12.57)} & \multicolumn{2}{c|}{72.99 (15.82)} & 74.89 \\
    CA-JDA & \multicolumn{2}{c|}{81.07 (11.19)} & \multicolumn{2}{c|}{74.15 (15.77)} & 77.61 \\
    CA-JGSA & \multicolumn{2}{c|}{76.79 (12.35)} & \multicolumn{2}{c|}{73.07 (16.33)} & 74.93 \\ \midrule
    MEKT-E & \multicolumn{2}{c|}{81.29 (10.18)} & \multicolumn{2}{c|}{76.00 (17.61)} & 78.65 \\
    MEKT-L & \multicolumn{2}{c|}{\underline{83.07} (9.30)} & \multicolumn{2}{c|}{\textbf{76.54} (16.72)} & \underline{79.81} \\
    MEKT-R & \multicolumn{2}{c|}{\textbf{83.42} (9.55)} & \multicolumn{2}{c|}{\underline{76.31} (16.76)} & \textbf{79.87} \\ \midrule
          & \multicolumn{2}{c|}{RSVP} & \multicolumn{2}{c|}{ERN} & Avg \\ \midrule
     CSP-LDA  & \multicolumn{2}{c|}{65.36 (9.32)} & \multicolumn{2}{c|}{61.87 (4.51)} & 63.62 \\
    EA-CSP-LDA & \multicolumn{2}{c|}{\textbf{69.07} (9.05)} & \multicolumn{2}{c|}{64.63 (5.86)} & 66.85 \\
     RA-MDM & \multicolumn{2}{c|}{67.29 (8.38)} & \multicolumn{2}{c|}{62.90 (6.79)} & 65.10 \\
    CA    & \multicolumn{2}{c|}{67.35 (7.52)} & \multicolumn{2}{c|}{65.89 (7.30)} & 66.62 \\
    CA-CORAL & \multicolumn{2}{c|}{66.94 (7.46)} & \multicolumn{2}{c|}{66.17 (7.74)} & 66.56 \\
    CA-GFK & \multicolumn{2}{c|}{67.75 (7.48)} & \multicolumn{2}{c|}{66.03 (7.50)} & 66.89 \\
    CA-JDA & \multicolumn{2}{c|}{66.06 (6.18)} & \multicolumn{2}{c|}{64.64 (6.50)} & 65.35 \\
    CA-JGSA & \multicolumn{2}{c|}{64.57 (5.79)} & \multicolumn{2}{c|}{57.68 (8.04)} & 61.13 \\ \midrule
    MEKT-E & \multicolumn{2}{c|}{67.92 (6.70)} & \multicolumn{2}{c|}{\textbf{66.70} (8.00)} & \textbf{67.31} \\
    MEKT-L & \multicolumn{2}{c|}{\underline{68.40} (6.40)} & \multicolumn{2}{c|}{65.98 (7.94)} & 67.19 \\
    MEKT-R & \multicolumn{2}{c|}{68.38 (6.36)} & \multicolumn{2}{c|}{\underline{66.17} (7.68)} & \underline{67.28} \\ \bottomrule
    \end{tabular}
  \label{tab:bca2}
\end{table}

\begin{figure}[htpb]  \centering
\includegraphics[width=\linewidth,clip]{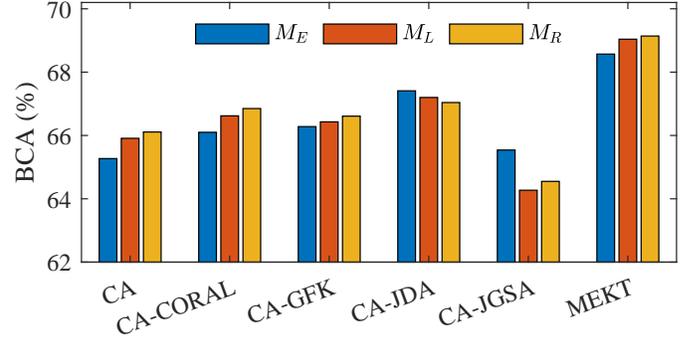}
\caption{Average BCAs (\%) of the tangent space approaches on the four datasets, when different reference matrices were used in CA.} \label{fig:ref}
\end{figure}

Fig.~\ref{fig:ref} shows the BCAs of all tangent space based approaches when different reference matrices were used in CA. The Riemannian mean obtained the best BCA in four out of the six approaches, and also the best overall performance.

We also performed paired $t$-tests on the BCAs to check if the performance improvements of MEKT-R over others were statistically significant. Before each $t$-test, we performed a Lilliefors test \cite{lilliefors1967} to verify that the null hypothesis that the data come from a normal distribution cannot be rejected. Then, we performed false discovery rate corrections \cite{benjamini1995} by a linear-step up procedure under a fixed significance level ($\alpha=0.05$) on the paired $p$-values of each task.

The false discovery rate adjusted $p$-values ($q$-values) are shown in Table~\ref{tab:psts}. MEKT-R significantly outperformed all baselines in almost all STS transfers. The performance improvements became less significant when there were multiple source domains, which is reasonable, because generally in machine learning the differences between different algorithms diminish as the amount of training data increases.

\begin{table}[htbp]  \centering \setlength{\tabcolsep}{2mm}
  \caption{False discovery rate adjusted $p$-values in paired $t$-tests ($\alpha=0.05$).}
    \begin{tabular}{c|l|cccc}  \toprule
    & \multicolumn{1}{c|}{MEKT-R vs}  & MI1   & MI2   & RSVP  & ERN \\  \midrule
\multirow{11}[1]{*}{STS}  &  CSP-LDA & \textbf{.0000} & \textbf{.0000} & --& --\\
  &  xDAWN-SVM & -- & -- & \textbf{.0002} & \textbf{.0000} \\
  &  EA-CSP-LDA & \textbf{.0030} & \textbf{.0003} &-- & --\\
  &  EA-xDAWN-SVM &-- &-- & \textbf{.0000} & \textbf{.0000} \\
  &  RA-MDM & \textbf{.0003} & \textbf{.0340} & \textbf{.0412} & \textbf{.0004} \\
  &  CA & \textbf{.0000} & \textbf{.0006} & \textbf{.0000} & \textbf{.0010} \\
  &  CA-CORAL & \textbf{.0005} & \textbf{.0340} & \textbf{.0000} & \textbf{.0014} \\
  &  CA-GFK & \textbf{.0000} & \textbf{.0001} & \textbf{.0016} & \textbf{.0130} \\
  &  CA-JDA & \textbf{.0003} & \textbf{.0183} & \textbf{.0386} & .2627 \\
  &  CA-JGSA & \textbf{.0021} & \textbf{.0006} & \textbf{.0000} & \textbf{.0241} \\  \midrule
 \multirow{11}[1]{*}{MTS}&     CSP-LDA & \textbf{.0329} & .1140 & --& --\\
 &   xDAWN-SVM &-- &-- & .2077 & \textbf{.0306} \\
 &   EA-CSP-LDA & .2808 & .1636 &-- &-- \\
 &   EA-xDAWN-SVM &-- &-- & .5733 & .2632 \\
 &   xDAWN-RA-MDM & .0824 & .1636 & .5347 & .0632 \\
 &   CA & \textbf{.0329} & .1260 &.4727 & .8380 \\
 &   CA-CORAL & .0897 & .1636 & .3477 & .9914 \\
 &   CA-GFK & .0824 & .1260 & .5347 & .9117 \\
 &   CA-JDA & .2379 & .1636 & \textbf{.0349} & .0632 \\
 &   CA-JGSA & .1344 & .1636 & \textbf{.0323} & \textbf{.0018} \\  \bottomrule
    \end{tabular}   \label{tab:psts}
\end{table}

We also considered linear and radial basis function (RBF; kernel width 0.1) kernels in MEKT-R, and repeated the above experiments. The results are shown in Table~\ref{tab:kernel}, where \emph{Primal} denotes the primal MEKT-R without kernelization. The primal MEKT-R achieved the best (in bold) or the second best (underlined) performance in all scenarios. However, the differences among the three approaches were very small.

\begin{table}[htbp]  \centering \setlength{\tabcolsep}{5mm}
  \caption{Average BCAs(\%) of the proposed MEKT under different kernels.}
    \begin{tabular}{c|c|ccc}  \toprule
     \multicolumn{1}{c}{}& & $\text{Primal}$ & $\text{Linear}$ & $\text{RBF}$ \\ \midrule
        \multirow{4}{*}{STS}
          & MI1   & \textbf{70.99} & \textbf{70.99} & 70.37 \\
          & MI2   & \textbf{68.73} & \textbf{68.73} & 68.37 \\
          & RSVP  & \underline{61.24} & 60.49 & \textbf{61.78} \\
          & ERN   & \underline{57.85} & 57.44 & \textbf{58.45} \\ \midrule
       \multirow{4}{*}{MTS}
          & MI1   & \textbf{83.42} & 83.36 & 78.21 \\
          & MI2   & \textbf{76.31} & \textbf{76.31} & 76.08 \\
          & RSVP  & \underline{68.38} & \textbf{68.41} & 68.22 \\
          & ERN   & \textbf{66.17} & 66.02 & 65.49 \\ \midrule
    \multicolumn{1}{c}{}& Avg & \textbf{69.14} & 68.97 & 68.37 \\  \bottomrule
    \end{tabular} \label{tab:kernel}
\end{table}

\subsection{Computational Cost}

This subsection empirically checked the computational cost of different algorithms, which were implemented in Matlab 2018a on a laptop with i7-8550U CPU@2.00GHz, 8GB memory, running 64-bit Windows 10 Education Edition.

For simplicity, we only selected one transfer task in each dataset. For STS transfer, the first subject in each dataset was selected as the target domain, and the second subject as the source domain. For MTS transfer, the first subject as the target domain, and all other subjects as the source domains. we repeated the experiment 20 times, and show the average computing time in Table~\ref{tab:time}. In summary, EA was the most efficient. RA-MDM, CA-JDA and MEKT-R had similar computational cost. MEKT-L and MEKT-E had comparable classification performance with MEKT-R (Tables~\ref{tab:bca1} and ~\ref{tab:bca2}), but much less computational cost. MEKT-L achieved the best compromise between the classification accuracy and the computational cost.

\begin{table}[htbp]  \centering \setlength{\tabcolsep}{1mm}
  \caption{Computing time (seconds) of different approaches in STS and MTS transfers.}
    \begin{tabular}{c|c|cccccc}  \toprule
     \multicolumn{1}{c}{}& & RA-MDM & EA & CA-JDA & MEKT-E & MEKT-L & MEKT-R \\ \midrule
        \multirow{4}{*}{STS}
          & MI1 & 5.49  & \textbf{0.44} & 5.45  & \underline{2.53}  & 2.75  & 5.42 \\
          & MI2 & 0.48  & \textbf{0.27} & 0.54  & \underline{0.43}  & 0.47  & 0.60 \\
          & RSVP& 0.42  & \textbf{0.05} & 0.45  & \underline{0.23}  & 0.27  & 0.43 \\
          & ERN & 0.54  & 0.47  & 0.53  & \textbf{0.38} & \underline{0.42}  & 0.53 \\ \midrule
       \multirow{4}{*}{MTS}
          & MI1 & 13.61 & \textbf{0.94} & \underline{9.24}  & 11.06 & 11.48 & 12.96 \\
          & MI2 & \underline{1.01}  & \textbf{0.69} & 1.35  & 1.13  & 1.20  & 1.29 \\
          & RSVP& \underline{3.13}  & \textbf{1.08} & 8.64  & 5.61  & 5.98  & 6.74 \\
          & ERN & \textbf{5.49} & 7.95  & 14.92 & \underline{10.39} & 10.74 & 11.95 \\ \bottomrule
    \end{tabular} \label{tab:time}
\end{table}

\subsection{Effectiveness of the Joint Probability MMD}

To validate the superiority of the joint probability MMD over the traditional MMD, we replaced the joint probability MMD term $\mathcal{D}'_{S,T}$ in (\ref{eq:MEKT}) by the traditional MMD term $\mathcal{D}_{S,T}$ in (\ref{eq:MMD}), and repeated the experiments. The results are shown in Table~\ref{tab:mmd}. The joint probability MMD outperformed the traditional MMD in six out of the eight tasks. We expect that the joint probability MMD should also be advantageous in other applications that the traditional MMD is now used.

\begin{table}[htbp]  \centering \setlength{\tabcolsep}{3mm}
  \caption{Average BCAs (\%) when $\mathcal{D}_{S,T}$ in (\ref{eq:MMD}) or $\mathcal{D}_{S,T}'$ in (\ref{eq:JPMMD}) was used in (\ref{eq:MEKT}).}
    \begin{tabular}{c|c|cc}  \toprule
     \multicolumn{1}{c}{}& & $\mathcal{D}_{S,T}$ & $\mathcal{D}_{S,T}'$  \\ \midrule
        \multirow{4}{*}{STS} & MI1 &65.33& \textbf{70.99}\\
          & MI2& 66.78& \textbf{68.73}\\
             & RSVP& 61.11& \textbf{61.24}\\
               & ERN & \textbf{58.62} & 57.85\\
       \multirow{4}{*}{MTS}          & MI1 & 73.86& \textbf{83.42}\\
         & MI2& 74.23& \textbf{76.31}\\
            & RSVP & \textbf{69.33}& 68.38\\
             & ERN & 65.59& \textbf{66.17}\\ \midrule
    \multicolumn{1}{c}{}& Avg& 66.86& \textbf{69.14}    \\  \bottomrule
    \end{tabular}    \label{tab:mmd}
\end{table}

\subsection{Effectiveness of DTE}

This subsection validates our DTE strategy on MTS tasks to select the most beneficial source subjects.

Table~\ref{tab:dte_bca} shows the BCAs when different source domain selection approaches were used: RAND randomly selected $round[(z-1)/2]$ source subjects [because there was randomness, we repeated the experiment 20 times, and report the mean and standard deviation (in the parentheses)], ROD was the approach proposed in  \cite{Gong2012}, and ALL used all $z$ source subjects. Table~\ref{tab:dte_time} shows the computational cost of different algorithms.

Tables~\ref{tab:dte_bca} and \ref{tab:dte_time} shows that the proposed DTE outperformed RAND and ROD in terms of the classification accuracy. Although its BCAs were generally slightly worse than those of ALL, its computational cost was much lower than ALL, especially when $z$ became large, i.e., when $z \gg 1$, it can save over 50\% computational cost.

\begin{table}[htbp]  \centering \setlength{\tabcolsep}{3mm}
  \caption{Average BCAs (\%) with different source domain selection approaches. RAND, ROD and DTE each selected $round[(z-1)/2]$ source subjects. ALL used all source subjects.}
    \begin{tabular}{c|c|cccc}  \toprule
          & $z$& RAND  & ROD   & DTE   & ALL \\  \midrule
    MI1   & 7  & 81.53 (1.19) & 81.86 & \underline{82.14} & \textbf{83.42} \\
    MI2   & 9  & 75.05 (1.06) & 74.38 & \underline{76.23} & \textbf{76.31} \\
    RSVP  & 11 & 67.48 (0.31)  & 67.79  & \textbf{68.70} & \underline{68.38} \\
    ERN   & 16 & 65.31 (0.52) & 65.36 & \underline{65.51} & \textbf{66.17} \\  \bottomrule
    \end{tabular}   \label{tab:dte_bca}
\end{table}

\begin{table}[htbp]  \centering \setlength{\tabcolsep}{3mm}
  \caption{Computing time (seconds) of different source domain selection approaches. RAND, ROD and DTE each selected $round[(z-1)/2]$ source subjects. ALL used all source subjects.}
    \begin{tabular}{c|c|cccc}  \toprule
          & $z$ & RAND  & ROD   & DTE   & ALL \\  \midrule
    MI1   & 7     & \textbf{11.55} & 12.46 & \underline{11.77} & 12.84 \\
    MI2   & 9     & \textbf{0.90}  & 1.11  & \underline{0.94}  & 1.24 \\
    RSVP  & 11    & \textbf{3.08}  & 3.22  & \underline{3.10}  & 6.80 \\
    ERN   & 16    & \textbf{6.27}  & 6.42  & \underline{6.29}  & 11.57 \\  \bottomrule
    \end{tabular}   \label{tab:dte_time}
\end{table}

\section{Conclusions}

Transfer learning is popular in EEG-based BCIs to cope with variations among different subjects and/or tasks. This paper has considered offline unsupervised cross-subject EEG classification, i.e., we have labeled EEG trials from one or more source subjects, but only unlabeled EEG trials from the target subject. We proposed a novel MEKT approach, which has three steps: 1) align the covariance matrices of the EEG trials in the Riemannian manifold; 2) extract tangent space features; and, 3) perform domain adaptation by minimizing the joint probability distribution shift between the source and the target domains, while preserving their geometric structures. An optional fourth step, DTE, was also proposed to identify the most beneficial source domains, and hence to reduce the computational cost. Experiments on four EEG datasets from two different BCI paradigms demonstrated that MEKT outperformed several state-of-the-art transfer learning approaches. Moreover, DTE can reduce more than half of the computational cost when the number of source subjects is large, with little sacrifice of classification accuracy.



\end{document}